\documentclass[11pt,epsf,letterpaper]{article}%
\usepackage{setspace}
\usepackage{amsmath}
\usepackage{color}
\usepackage{amsfonts}
\usepackage{verbatim}
\usepackage{amssymb}
\usepackage{graphicx}
\usepackage{amsmath}
\usepackage{amsfonts}
\usepackage{amssymb}
\usepackage{amsmath}%
\providecommand{\U}[1]{\protect\rule{.1in}{.1in}}

\def\b2hat{ {\hat b}_2 }

\def\D{\tilde{\nabla}}

\def\Cn{N^{(k)}_D}
\begin{document}

\title{Birkhoff's Theorem in Higher Derivative Theories of Gravity}
\author{Julio Oliva$^{1}$, Sourya Ray$^{2}$\\$^{1}$\textit{Instituto de F\'{\i}sica, Facultad de Ciencias, Universidad
Austral de Chile, Valdivia, Chile.}\\$^{2}$\textit{Centro de Estudios Cient\'{\i}ficos (CECS), Casilla 1469,
Valdivia, Chile.}\\{\small julio.oliva@docentes.uach.cl, ray@cecs.cl}}
\date{}
\maketitle
\begin{abstract}
In this paper we present a class of higher derivative theories of gravity
which admit Birkhoff's theorem. In particular, we explicitly show that in this
class of theories, although generically the field equations are of fourth
order, under spherical (plane or hyperbolic) symmetry, all the field
equations reduce to second order and have exactly the same or similar
structure to those of Lovelock theories, depending on the spacetime dimensions
and the order of the Lagrangian.

\end{abstract}

\pagebreak

\section{Introduction}

In General Relativity, Birkhoff's theorem states that the spherically
symmetric solutions of Einstein's equations in the vacuum are locally
isometric to the Schwarzschild solution. This theorem has recently been
generalized for a class of higher curvature theories namely the Lovelock
theories \cite{Zegers}. These theories are natural generalizations of
Einstein's theory in higher dimensions. While the field equations of a general
higher curvature theory involves second derivatives of the Riemann tensor,
Lovelock theories share the property of Einstein gravity that no derivative of
the curvature tensor arise and hence the field equations are second order in
the derivative of the metric \cite{lovelock}. In $D$-spacetime dimensions the general Lovelock Lagrangian consists of an arbitrary linear combination of all the
$2k$-dimensional Euler densities, where $2k<D$. Birkhoff's theorem in Lovelock
gravity states that spherically (plane or hyperbolic) symmetric solutions of
the Lovelock field equations are isometric to the corresponding static
Lovelock black hole solution. The proof relies on the second order nature of
the Lovelock field equations. The admittance of Birkhoff's theorem is related
to the lack of spin-$0$ mode excitations in the linearized field equations. It is of fundamental importance in proving uniqueness theorems in General Relativity.

Lately, there has been a great interest in other higher curvature theories of
gravity coming from high energy physics, cosmology and astrophysics. These
theories generically have higher order field equations in the metric
formalism. However, as of yet, none of these theories are known to admit
Birkhoff's theorem. In this article, we present a class of higher derivative theories which includes the Lovelock theories as a subclass and admits Birkhoff's theorem. These theories are characterized by the Lagrangian densities which give second order traced field equations and further all the field equations reduce to second order for spherically (plane or hyperbolic) symmetric ansatz.

\section{Construction of Lagrangian densities}

It is well known \cite{Far} that in dimensions greater or
equal to four, the most general Lagrangian density which is
quadratic in the curvature and gives second order traced field equations can
be expressed as an arbitrary linear combination of the Gauss-Bonnet density
and the quadratic conformal density
\footnote{Hereafter, by a conformal density we mean a scalar invariant constructed by
contracting all the indices of a certain number of conformal tensors among
themselves times $\sqrt{-g}$.}.
The Gauss-Bonnet density being the quadratic Lovelock Lagrangian gives second
order field equations for generic metrics. The field equations obtained from
the quadratic conformal density on the other hand is generically of fourth
order. However, the trace of the field equations is proportional to the
invariant itself, the proportionality factor being $(4-D)/2$. This is not very
surprising if one considers the variation of the action under infinitesimal
conformal rescalings of the metric $\delta g^{ab}=\Omega g^{ab}$ which gives
\begin{align}
\delta I=\int\sqrt{-g}\mathcal{E}_{ab}\delta g^{ab}=\int\delta(\sqrt
{-g}\mathcal{L})=\int\Omega\left(  2-\frac{D}{2}\right)  \sqrt{-g}\mathcal{L},%
\end{align}
where $\mathcal{L}=C_{ab}^{\ \ cd}C_{cd}^{\ \ ab}$. It turns out that in
dimensions greater or equal to four the Gauss-Bonnet density and the quadratic
conformal density are the only two linearly independent invariants which
generically gives second order traced field equations. Thus there is a two
dimensional space of quadratic invariants with this special property in
dimensions greater or equal to four. 

Now let us consider the same question in three dimensions. One might naively be tempted to think that since in three dimensions, the conformal tensor identically vanishes and so does the Gauss-Bonnet invariant, there are no quadratic invariants which gives second
order traced field equations. However, this assertion can be explicitly shown
to be incorrect. One may easily check that the density 
\begin{equation}
 \sqrt{-g}\left[R^{ab}R_{ab}-\frac{3}{8}R^{2}\right]
 \label{nmgaction}
\end{equation}
does give field equations whose trace is of second order
and is proportional to the density itself \footnote{This density supplemented by an Einstein-Hilbert term and a cosmological constant has lately gained a lot of attention in the community as a toy model for quantum gravity in three dimensions and is known as the theory of New Massive Gravity \cite{bht2009}}. In fact this is the only
quadratic invariant which has this special property in three dimensions. Let
us now ask ourselves the same question for cubic densities. First let us try
to guess all the linearly independent cubic invariants which has this
property. Obviously, first there is the six-dimensional Euler density all of
whose field equations are of second order. Next one can repeat the same
argument for the quadratic conformal density to the cubic conformal densities.
Note that in dimensions greater than five there are two independent ways of
contracting three conformal tensors. Hence, in dimensions greater or equal to
five, there are at least three independent invariants which give second order
traced field equations, namely- the six-dimensional Euler density and the two independent cubic
conformal densities. In \cite{or2}, we had explicitly shown that these are the
only invariants which possess this property. However, in lower dimensions the
Euler density identically vanishes and the two conformal densities are not
linearly independent. In fact, in four dimensions the only cubic invariant
which has the property is given by the conformal density. Whereas in five
dimensions in addition to the cubic conformal density there is another
independent density which shares the property. It is given by
\begin{align}
\sqrt{-g} \left[24R_{ab}^{\ \ cd}R_{cd}^{\ \ be}R_{e}^{\ a}+\frac{21}{4}R_{ab}^{\ \ cd}R_{cd}%
^{\ \ ab}R+40R_{ab}^{\ \ cd}R_{a}^{\ c}R_{b}^{\ d}+\frac{320}{9}R_{a}^{\ b}R_{b}^{\ c}%
R_{c}^{\ a}-\frac{97}{3}R_{a}^{\ b}R_{b}^{\ a}R+\frac{31}{9}R^{3}\right].%
\label{splcubicinv}
\end{align}
This invariant is the cubic counterpart of the special quadratic density (\ref{nmgaction}) in
three dimensions. Now, to generalize the case for arbitrary higher order one
needs to classify or understand the nature of these special invariants.

Let us scrutinize the special quadratic invariant more carefully. First
realize that in dimensions four and higher there are three linearly
independent Riemannian invariants namely - $R^{abcd}R_{abcd},R^{ab}R_{ab}$ and
$R^{2}$. However, in three dimensions only two of them are independent. In
other words any one of the three invariants can be expressed in terms of the
other two. This identity is responsible for vanishing of both the Gauss Bonnet
density and the conformal density \footnote{Of course this can be equivalently
understood in terms of vanishing of the conformal tensor. However, here we are
concerned with scalar identities.}. Analogously, there are eight linearly
independent cubic Riemannian invariants in dimensions greater or equal to six.
However, in five (and lower) dimensions the six-dimensional Euler density
vanishes identically. In fact, this is the only independent identity in five
dimensions among the cubic scalar invariants. So, this identity is also
responsible for the linear dependence of the two conformal invariants. The
same occurs for invariants of arbitrary higher order $k$ in dimensions
$D=2k-1$ because of the following relation
\begin{align}
\delta_{a_{1}b_{1}\cdots a_{k}b_{k}}^{c_{1}d_{1}\cdots c_{k}d_{k}}%
R_{c_{1}d_{1}}^{\ \ \ a_{1}b_{1}}\cdots R_{c_{k}d_{k}}^{\ \ \ a_{k}b_{k}}=\delta
_{a_{1}b_{1}\cdots a_{k}b_{k}}^{c_{1}d_{1}\cdots c_{k}d_{k}}C_{c_{1}d_{1}%
}^{\ \ \ a_{1}b_{1}}\cdots C_{c_{k}d_{k}}^{\ \ \ a_{k}b_{k}}=0,
\label{identity}
\end{align}
where $\delta^{\cdots}_{\cdots}$ is the generalized Kronecker delta. Now, in the case of quadratic invariants, when one takes a particular linear combination of the Gauss-Bonnet density and the conformal density in arbitrary dimensions and reexpresses the conformal invariant in terms of the Riemannian invariants then it factorizes by $(D-3)$. Explicitly,
\begin{align}
R^{abcd}R_{abcd}-4R^{ab}R_{ab}+R^{2}-C^{abcd}C_{abcd}=-\frac{D-3}{D-2}%
\left(4R^{ab}R_{ab}-\frac{D}{D-1}R^{2}\right).
\end{align}
Obviously, since the left hand side is a linear combination of the
Gauss-Bonnet density and the conformal density, it gives second order traced
field equations in arbitrary dimensions. This implies that the invariant
inside the parenthesis on the right-hand side also gives second order traced
field equation in arbitrary dimensions. However, the left-hand side vanishes
in three dimensions whereas the term inside the parenthesis on the right-hand
side does not. In fact in three dimensions it gives the special invariant
mentioned earlier. Similar situation arises in the cubic case where a
particular linear combination of the six-dimensional Euler density and the two
conformal invariants can be factorized by $D-5$ and the remaining invariant
does not vanish in five dimensions identically. This gives the special cubic
invariant in five dimensions (\ref{splcubicinv}). Interestingly this can be generalized to
arbitrary higher orders when one realizes that there is always a particular
combination of the $2k$-dimensional Euler density and the $k$-th order
conformal invariants which can be expressed as
\begin{align}
\delta_{a_{1}b_{1}\cdots a_{k}b_{k}}^{c_{1}d_{1}\cdots c_{k}d_{k}}%
(R_{c_{1}d_{1}}^{\ \ \ a_{1}b_{1}}\cdots R_{c_{k}d_{k}}^{\ \ \ a_{k}b_{k}}-C_{c_{1}d_{1}%
}^{\ \ \ a_{1}b_{1}}\cdots C_{c_{k}d_{k}}^{\ \ \ a_{k}b_{k}}),
\end{align}
which when expanded (reexpressed) in terms of Riemannian invariants is
factorized by $(D-2k+1)$. The remaining invariant is non-vanishing in
dimensions $D=2k-1$ but vanishes identically in lower dimensions. Based on
this observation we had proposed a conjecture in \cite{or2} that any $k$th
order Riemannian invariant which gives second order traced field equations can
be expressed as $\sqrt{-g}$ times a linear combination of the special invariant, all the
linearly independent conformal invariants and a divergence term \footnote{We have checked this conjecture up to order four.}. Explicitly, if in $D$ dimensions, there are $\Cn$ linearly independent $k(\geq2)$th order conformal invariants $W_{i}^{(k)}$ for \{$i=1,\cdots \Cn$\}, then the most general action which gives
second order traced field equations is given by 
\begin{align}
I^{(k)}&=\int\sqrt{-g}\left(\alpha^{(k)}_{0}\mathcal{N}^{(k)}+\sum_{i=1}^{\Cn}%
\alpha^{(k)}_{i}W_{i}^{(k)}+\text{a divergence term}\right),
\label{actionW}\\
\text{where} \qquad \mathcal{N}^{(k)}&=\frac{1}{2^{k}}\left(  \frac{D-2}{D-2k+1}\right)
\delta_{a_{1}b_{1}\cdots a_{k}b_{k}}^{c_{1}d_{1}\cdots c_{k}d_{k}}%
(R_{c_{1}d_{1}}^{\ \ \ a_{1}b_{1}}\cdots R_{c_{k}d_{k}}^{\ \ \ a_{k}b_{k}}-C_{c_{1}d_{1}%
}^{\ \ \ a_{1}b_{1}}\cdots C_{c_{k}d_{k}}^{\ \ \ a_{k}b_{k}}).
\label{spclinvariant}
\end{align}
Note that in dimensions $D\geq2k$, the special invariant $\mathcal{N}^{(k)}$
can be expressed as a linear combination of the $2k$-dimensional Euler density
and all the conformal invariants and in dimensions $D<2k-1$, it vanishes identically.

Now we show that a subclass of these theories which generically gives fourth
order field equations but {\it for spherically (plane or hyperbolic) symmetric spacetimes all the field
equations reduce to second order!}. And it is in this class where Birkhoff's
theorem can be shown to hold.

\section{Field equations for spherically (plane or hyperbolic) symmetric spacetimes}

Consider the general spherically (plane or hyperbolic) symmetric spacetimes
given by the following line element
\begin{align}
ds^{2}=\tilde{g}_{ij}(x)dx^{i}dx^{j}+e^{2\lambda(x)}d\Sigma_{\gamma}^{2},%
\label{ansatz}
\end{align}
where $d\Sigma_{\gamma}^{2}=\hat{g}_{\alpha\beta}(y)dy^{\alpha}dy^{\beta}$ is
the line element of a $(D-2)$-dimensional space of constant curvature $\gamma
$. Let $\D$ be the Levi-Civita connection on the two-dimensional space orthogonal to the constant curvature space and $\tilde{R}$ be the corresponding scalar curvature. Then the nontrivial components of the Riemann curvature tensor and the conformal tensor are given by
\begin{align}
&  R_{jl}^{\ \ ik}=\frac{1}{2}\tilde{R}\delta_{jl}^{ik},\qquad \qquad \qquad \qquad R_{\nu\rho}%
^{\ \ \mu\lambda}=\tilde{\mathcal{B}}\delta_{\nu\rho}^{\mu\lambda},\qquad \qquad \qquad \qquad R_{j\nu}^{\ \ i\mu}=-\tilde{\mathcal{A}}_{j}^{i}\delta_{\nu
}^{\mu},&\\
&  C_{jl}^{\ \ ik}=\frac{(D-3)\tilde{S}}{2(D-1)}\delta_{jl}^{ik},\qquad C_{\nu\rho
}^{\ \ \mu\lambda}=\frac{\tilde{S}}{(D-1)(D-2)}\delta_{\nu\rho}^{\mu\lambda},\qquad
C_{j\nu}^{\ \ i\mu}=-\frac{(D-3)\tilde{S}}{2(D-1)(D-2)}\delta_{j}^{i}\delta_{\nu
}^{\mu},&
\end{align}
where 
\begin{align}
 &\tilde{\mathcal{B}}=\gamma e^{-2\lambda}-(\D_{m}\lambda
)(\D^{m}\lambda), \\
&\tilde{\mathcal{A}}_{j}^{i}=\D^{i}%
\D_{j}\lambda+(\D^{i}\lambda)(\D_{j}\lambda), \\
\text{and}\ \ \ &\tilde{S}=\tilde{R}+2\D_{k}\D^{k}\lambda+2\gamma
e^{-2\lambda}.
\end{align}
%
Note that since all the components of the conformal tensor are
a mere multiple of the function $\tilde{S}$, each of the conformal densities
$W_{i}^{(k)}$'s evaluated on the metic (\ref{ansatz}) are proportional to
${\tilde{S}}^{k}$. Let $W_{m}^{(k)}=\omega_{m}(D,k){\tilde{S}}^{k}$. Then the
field equations for the action (\ref{actionW}) evaluated on the metric ansatz (\ref{ansatz}) are given by
\begin{align}
\mathcal{G}_{\ \ \ j}^{(k)i} &  =\frac{(D-2)!(D-2)\alpha^{(k)}_{0}\tilde{\mathcal{B}}^{k-1}%
}{2(D-2k-1)!(D-2k+1)}\left[  2k\delta_{jl}^{ik}\tilde{\mathcal{A}}_{k}%
^{l}-(D-2k-1)\delta_{j}^{i}\tilde{\mathcal{B}})\right]  \nonumber\\
&  +k\left(  \sum_{m=1}^{\Cn}\alpha^{(k)}_{m}\omega_{m}(D,k)-\frac{(D-2)\alpha^{(k)}
_{0}\omega_{0}(D,k)}{2^{k}(D-2k+1)}\right)  \tilde{\mathcal{P}}_{j}^{i}%
(\tilde{S}^{k-1}),
\label{fe1}
\end{align}
\begin{align}
\mathcal{G}_{\ \ \ \beta}^{(k)\alpha} &  =-\frac{(D-2)!\alpha^{(k)}_{0}\tilde{\mathcal{B}%
}^{k-2}}{(D-2k-1)!(D-2k+1)}{\delta}_{\beta}^{\alpha}%
\Bigl[(D-2k-1)(D-2k-2)\tilde{\mathcal{B}}^{2}\nonumber\\
&  +k(\tilde{R}-2(D-2k-1){\tilde{\mathcal{A}}}_{i}^{i})\tilde{\mathcal{B}%
}+2k(k-1)\delta_{jl}^{ik}{\tilde{\mathcal{A}}}_{i}^{j}{\tilde{\mathcal{A}}%
}_{k}^{l}\Bigr]\nonumber\\
&  +k\left(  \sum_{m=1}^{\Cn}\alpha^{(k)}_{m}\omega_{m}(D,k)-\frac{(D-2)\alpha^{(k)}
_{0}\omega_{0}(D,k)}{2^{k}(D-2k+1)}\right)  {\delta}_{\beta}^{\alpha}%
\tilde{\mathcal{Q}}(\tilde{S}^{k-1}),
\label{fe2}\\
\mathcal{G}_{\ \ \ \alpha}^{(k)i} &  =\mathcal{G}_{\ \ \ i}^{(k)\alpha}=0,
\end{align}
where $\tilde{\mathcal{P}}_{j}^{i}$ and $\tilde{\mathcal{Q}}$ are two (related)
second order linear differential operators defined on the two-dimensional
space orthogonal to the constant curvature base manifold and $\omega_0(D,k)$ is a positive number (see Appendix). Notice that the fourth derivative terms arise when the operators $\tilde{\mathcal{P}%
}_{j}^{i}$ and $\tilde{\mathcal{Q}}$ act on the function $\tilde{S}^{k-1}$.
However, in all the field equations these terms are multiplied by a numerical
factor which depends on the coupling constants $\alpha^{(k)}_{i}$, for \{$i=0,\cdots
\Cn$\}, the dimensions $D$ and the order $k$. Hence, if one chooses the coupling
constants in such a way that this factor vanishes then the field equations
reduce to second order. In particular, choosing \{$\alpha^{(k)}_{0}\neq0,\alpha^{(k)}_{i}%
$\} such that
\begin{align}
\sum_{m=1}^{\Cn}\alpha^{(k)}_{m}\omega_{m}(D,k)=\frac{(D-2)\alpha^{(k)}_{0}\omega_{0}%
(D,k)}{2^{k}(D-2k+1)},
\label{cond}
\end{align}
give the same equations as in pure Lovelock gravity theories in dimensions $D>2k$.
Note that for any dimension $D\geq2k$ for a given $k$, $\Cn=N^{(k)}_{2k}$ whereas $N^{(k)}_{2k}=N^{(k)}_{2k-1}+1$ as explained previously due to the identity (\ref{identity}).
Therefore, the number of independent densities of
order $k\geq2$ satisfying (\ref{cond}) in dimensions $D\geq2k$ is $N^{(k)}_{2k}$. Also, note that in $D=2k$ or $D<2k-1$, all the field equations corresponding to the densities satisfying (\ref{cond}) vanish identically when evaluated on the metric (\ref{ansatz}).

\section{Classification of theories admitting Birkhoff's theorem}

Let us now discuss the different cases up to the first
few orders. Since there are no conformal invariants of order one, the only
non-trivial action, in any dimension $D>2$, is given by the Einstein-Hilbert term. Next, there is only
one conformal invariant of order $k=2$ in dimensions $D\geq4$. This indicates
that there is only a one-parameter family of densities in $D>4$ which
admit Birkhoff's theorem. So, this must be the Gauss-Bonnet density. One
can explicitly check that this is indeed the case by calculating $\omega
_{0}(D,2)$ and $\omega_{1}(D,2)$. In $D=4$, the Gauss-Bonnet density is a topological term and in $D<4$ it vanishes identically and hence there are no densities quadratic in curvature which admits Birkhoff's theorem in dimensions $D\leq4$ \footnote{Interestingly, a weaker version of Birkhoff's theorem has been shown to hold for conformal gravity in four dimensions \cite{Riegert:1984zz}. There it has been proved that the most general spherically symmetric solution of conformal gravity is static {\it up to a conformal gauge transformation}.}. Now, let us consider the cubic invariants. It turns out that there are two independent conformal invariants in dimensions
$D\geq6$. Therefore, there must be two linearly independent densities which admit
Birkhoff's theorem in $D>6$. Obviously, a particular linear combination of these two densities gives the third order Euler density, which in turn has second order field equations for any metric and is already known to admit Birkhoff's theorem. This implies that any other linearly independent combination of the two densities represent a four-derivative theory, whose field equations when evaluated on the metric (\ref{ansatz}) are either the same as that of cubic Lovelock theory or are identically vanishing. Whereas in $D=6$, since the third order Euler density is a topological term, there is only one linearly independent non-trivial density which represents a four-derivative theory. However, when evaluated on the metric (\ref{ansatz}), all the corresponding field equations vanish identically. Now, in five dimensions, as explained previously, there is only
one independent cubic conformal invariant. Also, the third order Euler density vanishes identically in $D=5$. So, there is a unique cubic theory in $D=5$ which is a four-derivative theory and admits Birkhoff's theorem. Even though there is no cubic Lovelock theory in five dimensions, the field equations of this theory when evaluated on the metric (\ref{ansatz}) has a similar structure as that of the cubic Lovelock theory in higher dimensions. This
theory was first presented in \cite{or1} and the Birkhoff's theorem was proven
explicitly \footnote{It was simultaneously found in \cite{mr2010} where it was
named Quasi-topological gravity.}.

So for arbitrary order $k\geq2$, in dimensions 

$\mathbf{D>2k}$: there are $N^{(k)}_{2k}$ independent densities whose field equations are generically of fourth order but when evaluated on the metric (\ref{ansatz}) either give the same field equations as those of the corresponding $k$th order pure Lovelock theory or vanish identically. 

$\mathbf{D=2k}$: there are $N^{(k)}_{2k}$ independent densities out of which one particular linear combination gives the $2k$-dimensional Euler density which is a topological term. The field equations from any other linearly independent combination are generically of fourth order but when evaluated on the metric (\ref{ansatz}) vanish identically.

$\mathbf{D<2k-1}$: there are $\Cn-1$ independent densities whose field equations are generically of fourth order but when evaluated on the metric (\ref{ansatz}) vanish identically. This is because both the special invariant (\ref{spclinvariant}) and the right hand side of the equation (\ref{cond}) vanish. 

$\mathbf{D=2k-1}$: there are $N^{(k)}_{2k-1}=N^{(k)}_{2k}-1$ independent densities whose field equations are generically of fourth order but when evaluated on the metric (\ref{ansatz}) either vanish identically or reduce to second order and has a similar structural form as that of pure $k$th order Lovelock theory. In the later case the Birkhoff's theorem can be proved in the following way.

Let us begin by fixing coordinates in the two-dimensional spacetime with metric $\tilde{g}_{ij}$, and the
gauge freedom by choosing $\tilde{\nabla}e^{\lambda (x)}$ to be a
spacelike\footnote{%
The proof of the Birkhoff's theorem for timelike and null cases follows along the same lines.} vector on $\tilde{g}_{ij}$: 
\begin{equation}
ds^{2}=-f(t,r)dt^{2}+\frac{dr^{2}}{g(t,r)}+r^{2}d\Sigma_{\gamma}.
\label{timeanz}
\end{equation}%
The equation $\mathcal{G}_{\ \ \ t}^{(k)t}=0$ implies%
\begin{equation}
\left( \gamma -g\right) ^{k-1}\left( 2\left( \gamma -g\right) +krg^{\prime
}\right) =0\ ,  \label{tt}
\end{equation}%
while $\mathcal{G}_{\ \ \ r}^{(k)r}=0$ reduces to%
\begin{equation}
\left( \gamma -g\right) ^{k-1}\left( 2\left( \gamma -g\right) f+krgf^{\prime
}\right) =0\ ,  \label{rr}
\end{equation}
where the prime on $g$ and $f$ denote a partial differentiation with respect to $r$. Focussing first in the non-degenerate case, i.e. $g\left( t,r\right) \neq
\gamma $, from equation (\ref{tt}) we obtain%
\begin{equation}
g\left( t,r\right) =F_{1}\left( t\right) r^{2/k}+\gamma \ ,
\end{equation}%
where $F_{1}\left( t\right) $ is an arbitrary function of $t$. Replacing this in
equation (\ref{rr}), we obtain%
\begin{equation}
f\left( t,r\right) =F_{2}\left( t\right) g\left( t,r\right) \ .
\end{equation}%
The arbitrary function $F_{2}\left( t\right) $ can be reabsorbed by a
coordinate transformation without any loss of
generality. Consequently %
\begin{equation}
f\left( t,r\right) =g\left( t,r\right) =F_{1}\left( t\right) r^{2/k}+\gamma
\ .
\end{equation}%
Equation $\mathcal{G}_{\ \ \ r}^{(k)t}=0$ then implies $dF_{1}/dt =0
$, which in turn implies that $F_{1}$ is a constant $c$. Then the equations
along the base manifold $\mathcal{G}_{\ \ \ \beta }^{(k)\alpha }=0$ are satisfied
without any further restriction, and we obtain the following metric 
\begin{equation}
ds^{2}=-\left(  cr^{2/k}+\gamma\right)  dt^{2}+\frac{dr^{2}}{cr^{2/k}+\gamma
}+r^{2}d\Sigma_{\gamma}^{2}\ ,
\end{equation}
where $c$ is an integration constant. The spacetime is asymptotically locally
flat. In the hyperbolic case ($\gamma=-1$) the metric describes a black hole,
provided $c$ is positive. This completes the proof of the Birkhoff's Theorem for the non-degenerate case.

In the degenerate case where $g\left( t,r\right) =\gamma $, all the equations are trivially satisfied,
and the metric function $f\left( t,r\right) $, is left undetermined. Birkhoff's
theorem does not hold in this case. 

%
%

One can also consider actions which are non-homogenous in the order $k$.  In
such a case, unless there is a non-vanishing contribution to the action from
the invariant (\ref{spclinvariant}) of order $k=\tfrac{D+1}{2}$, the
spherically (plane or hyperbolic) symmetric solutions are given by the
corresponding solutions of general Lovelock theory of order $k=\left[
\tfrac{D-1}{2}\right] $. However, if there is such a contribution (necessarily
in odd dimensions) then the general spherically (plane or hyperbolic)
symmetric solution is again static but does not belong to the family of
solutions of general Lovelock theory. Nevertheless, the general solution has
the same structural form as that in general Lovelock theories of order
$k=\tfrac{D+1}{2}$ and can be implicitly written in terms of an algebraic
equation for the metric function $g_{tt}$. Generically, the solution is of the form
\begin{equation}
ds^{2}=-f\left( r\right) dt^{2}+\frac{dr^{2}}{f\left( r\right) }%
+r^{2}d\Sigma _{\gamma}^{2}\ ,
\end{equation}
where the function $f\left( r\right) $, solves the following polynomial
equation%
\begin{equation}
\sum_{k=0}^{\left[ \frac{D+1}{2}\right] }\tilde{\alpha}_{0}^{(k)}\left(
D-2k\right) r^{D-2k-1}\left( \gamma -f\left( r\right) \right) ^{k}=M\ ,
\end{equation}%
where $M$ is an integration constant and%
\begin{equation}
\tilde{\alpha}_{0}^{(k)}=\frac{\left( D-2\right) \left( D-2\right) !}{\left(
D-2k+1\right) !}\alpha _{0}^{(k)}\ .
\end{equation}

Therefore, in any dimension $D$, in addition to the cosmological constant, there is a $p$-parameter family of (non-trivial) Lagrangian densities which generically admit Birkhoff's theorem, where
\begin{align*}
 p =& 1+\left(\displaystyle\sum_{1<k< \tfrac{D}{2}}N^{(k)}_{2k}\right)+\left(\displaystyle\sum_{k\geq \tfrac{D}{2}}(N^{(k)}_D-1)\right)\ \ \ \text{for even} \ D\\
 & 1+\left(\displaystyle\sum_{1<k\leq \tfrac{D+1}{2}}N^{(k)}_{2k}-1\right)+\left(\displaystyle\sum_{k> \tfrac{D+1}{2}}(N^{(k)}_D-1)\right)\ \ \ \text{for odd} \ D.
\end{align*}
%
Here, for both even and odd $D$, the unity on the right hand side corresponds to the Einstein-Hilbert term, whereas the second and the third term correspond to the terms in the action which generically gives fourth order field equations but when evaluated on the metric ansatz (\ref{ansatz}) respectively reduces to second order and vanish entirely.

\section{Conclusions}

Our findings raise several natural questions. Firstly, is there an even wider
class of theories which admit Birkhoff's theorem? Secondly, we have seen that
the solutions of the theories discussed here all have the same structural form
namely those of Lovelock gravity theories. Is there any theory which admits
Birkhoff's theorem but the corresponding solution does not have the same
structure? It seems to us that, if it is necessary for Birkhoff's theorem to
hold, that the field equations reduce to second order under spherical (plane
or hyperbolic) symmetry then the corresponding solutions will have the same
structure.
\section*{Acknowledgements}
We thank Gast\'{o}n Giribet and Hideki Maeda for useful discussions and valuable comments. Special thanks to Francisco Correa and Jorge Zanelli for encouragement and support. This work is partially supported by the FONDECYT grant 11090281 and the CONICYT grant \textquotedblleft Southern Theoretical Physics Laboratory" ACT-91 grant. The Centro de Estudios Cient\'{\i}ficos (CECS) is funded by the Chilean Government through the Centers of Excellence Base Financing Program of CONICYT. 
\appendix
\section*{Appendix}
The operators $\tilde{\mathcal{P}}^{i}_{j}$ and $\tilde{\mathcal{Q}}$ in (\ref{fe1}) and (\ref{fe2}) are second order linear differential operators defined on the two-dimensional space orthogonal to the constant curvature base manifold and are given by 
\begin{align}
\tilde{\mathcal{P}}^{i}_{j}  &  =\Biggl[\delta^{i}_{j}\left(  \frac{\tilde{R}%
}{2}+(D-1)\D_{k}\D^{k}\lambda+(D-2)(D-1)\D_{k}%
\lambda\D^{k}\lambda+\D_{k}\D^{k}+(2D-3)\D%
_{k}\lambda\D^{k}-\frac{\tilde{S}}{2k}\right) \nonumber\\
&  -(D-2)(\D^{i}\D_{j}\lambda+D\D^{i}\lambda\D%
_{j}\lambda)-\D^{i}\D_{j}-(D-1)(\D^{i}\lambda\D_{j}+\D_{j}\lambda \D^{i}) \Biggr],\\
\tilde{\mathcal{Q}} & =-\frac{1}{D-2}\left[  \tilde{\mathcal{P}}^{i}_{i}%
-\tilde{S}\left(1-\frac{D}{2k}\right)\right].
\end{align}
The numerical factor $\omega_0(D,k)$ is given by
\begin{align}
\omega_{0}(D,k)=\frac{(D-2)!2^{k}}{(D-2k)!(D-2)^{k}(D-1)^{k}}\left[
(D-2k)(D-2k-1)+k(k-2)D(D-3)+k(k+1)(D-3)\right].
\end{align}

\end{document}